\documentclass[12pt,useAMS]{article}
\usepackage{natbib}
\usepackage{color}
\usepackage{mathrsfs}
\usepackage{amssymb,bm,amsmath,physics,amsthm}
\usepackage[colorlinks,linkcolor=blue,citecolor=blue,bookmarks=false]{hyperref}
\usepackage{multirow}
\usepackage{algorithm}
\usepackage{rotating}
\usepackage{algpseudocode}
\usepackage{graphicx}
\usepackage{tablefootnote}

\newcommand{\hbtheta}{\hat{\bm{\theta}}}

\def\eqx"#1"{{\label{#1}}}
\def\eqn"#1"{{\ref{#1}}}
\newcommand{\bt}[1]{\begin{center}\begin{tabular}{#1}\hline}
\newcommand{\et}{\\\hline\end{tabular}\end{center}}

\renewcommand{\hat}{\widehat}
\renewcommand{\bar}{\overline}
\newcommand{\bbeta}{{\mbox{\boldmath $\beta$}}}
\newcommand{\btheta}{{\mbox{\boldmath $\theta$}}}

\newcommand{\blambda}{{\mbox{\boldmath $\lambda$}}}

\newcommand{\1}{{\bf{{1}}}}

\usepackage{booktabs}
\usepackage[margin=1in, lines=25]{geometry}
\usepackage{setspace}
\newtheorem{proposition}{Proposition}
\newtheorem{definition}{Definition}
\newtheorem{remark}{Remark}
\renewcommand{\hat}{\widehat}
\begin{document}

\title{Bayesian Spatial
Homogeneity Pursuit for Survival Data with an Application to the SEER
Respiratory Cancer Data}
\author{Lijiang Geng\\
	   Department of Statistics, University of Connecticut, CT 06269, USA
	   \and 
	  Guanyu Hu\\
	   Department of Statistics, University of Missouri - Columbia, MO 65211, USA
	   }

\maketitle

\begin{abstract}
In this work, we propose a new Bayesian spatial homogeneity pursuit method 
for survival data under the proportional hazards model
to detect spatially clustered patterns in baseline hazard and
regression coefficients. Specially, regression coefficients and baseline 
hazard are assumed to have spatial homogeneity pattern over space. To capture
such homogeneity, we develop a
geographically weighted Chinese restaurant process prior to simultaneously 
estimate  coefficients and baseline hazards and their uncertainty measures. 
An efficient Markov chain Monte Carlo (MCMC) algorithm is designed
for our proposed methods. Performance is evaluated using simulated data, and
further applied to a real data analysis of respiratory cancer
in the state of Louisiana. 
\end{abstract}


\section{Introduction}\label{sec:intro}

Clinical data on individuals are often collected from different geographical
regions and then aggregated and analyzed in public health studies. The most
popular dataset is the Surveillance, Epidemiology, and End Results (SEER)
program \citep{2013seer} data which routinely collects population-based cancer
patient data from 20 registries across the United States. This data provides
prognostic and demographic factors of cancer patients. In this paper, we focus
our study on Louisiana respiratory cancer data which was analyzed in
\citet{mu2020bayesian}. Analysis of
such data conducted on a higher level often assume that covariate effects are
constant over the entire spatial domain. This is a rather
strong assumption, as all intrinsic heterogeneities in data are ignored. For
example, if one was to study the hazard for patients with lung cancer, it is
expected that the true hazard is not the same in areas where there is little air
pollution and severe air pollution, even for patients with similar
characteristics. From Tobler's first law of geography 
\citep{tobler1970computer}, it is reasonable to consider similarities between
nearby locations in
survival data due
to environmental circumstances in geographically close regions. In this paper,
we will recover the spatial homogeneity pattern of respiratory cancer survival
rates among different counties in state of Louisiana.

Existing 
approaches that account for such patterns in survival data can be put into 
two major categories. The first one is to incorporate spatial random effects in
survival models such as the accelerated failure time (AFT) model and the
proportional hazards
model
\citep{banerjee2003frailty,banerjee2005semiparametric,zhou2008joint,zhang2011bayesian,henderson2012modeling}, such that spatial variations are
accounted for by different
intercepts for different regions, while parameters for covariates are held 
constant. 
Another important approach, instead of assuming all covariate effects are
constant, allows parameters to be spatially varying in parametric,
nonparametric, and semiparametric models
\citep{hu2017spatial,hu2018modified,xue2019geographically}.

Despite their flexibility, the aforementioned spatially varying coefficients
models can be unnecessarily large. Imposing certain constraints on nearby
regions so that they have the same parameter values provides an efficient way of
reducing the model size without sacrificing too much of its flexibility. While
similar endeavor have been made to cluster spatial survival responses
\citep{huang2007detection, bhatt2014spatial}, the clustering of covariate
effects and baseline hazards have yet to be studied for survival data.

Two challenges are to be tackled for clustering of coefficients and baseline
hazards for spatial
survival models. First, the spatial structure needs to be appropriately
incorporated into the clustering process. Contiguousness constraints should be
added so that truly similar neighbors are driven to the same cluster. The
constraints, however, should not be overly emphasized, as two distant regions
may still share similar geographical and demographical characteristics and thus
parameters. Existing methods, such as in \cite{lee2017cluster,leespatial2019}
and \cite{li2019spatial}, do not allow for globally discontiguous clusters,
which is
a serious limitation. Second, the true number of clusters is unknown, and needs
to be estimated. With the probabilistic Bayesian framework, simultaneous
estimation of the number of clusters and the clustering configuration for each
region is achieved by complicated search algorithms
\citep[e.g., reversible jump MCMC,][]{green1995reversible}
in variable dimensional parameter spaces. Such algorithms assign a prior to the
number of clusters that needs to be updated in every MCMC iteration, which made
them difficult to implement or automate,
and suffer from mixing issues as
well as lack of scalability. Nonparametric Bayesian approaches, such as the
Chinese restaurant process \citep[CRP;][]{pitman1995exchangeable}, provide
another approach to allow for uncertainties in the number of clusters. Its
extension, the distance dependent CRP \cite[ddCRP;][]{blei2011distance},
considers spatial information, and makes a flexible class of distributions over
partitions that allows for dependencies between their elements. The CRP
framework, however, has been shown to be inconsistent in its estimation of
number of clusters \citep{miller2013simple}. \cite{lu2018reducing} proposed the
powered CRP that suppresses the small tail clusters. Similar to the traditional
CRP, however, it does not consider distance information, and therefore is not
well-suited when spatial homogeneity is to be detected.

To address these challenges, in this work, we consider a spatial proportional
hazards model, and propose a geographically weighted Chinese restaurant process
(gwCRP) to capture the spatial homogeneity of both the regression coefficients
and baseline hazards over subareas under piecewise constant hazards models
framework \citep{friedman1982piecewise}. Our main contributions in this paper
are three folds. First, we develop a new nonparametric Bayesian method for
spatial clustering which combines the ideas of geographical weights and
Dirichlet mixture models to leverage geographical information. Compared with
existing methods, our proposed approach is able to capture both locally
spatially contiguous clusters and globally discontiguous clusters. Second, an
efficient Markov chain Monte Carlo (MCMC) algorithm is proposed for our proposed
model without reversible jumps to simultaneously estimate the number of clusters
and clustering configuration.
In addition, we apply our method to the analysis of the Surveillance,
Epidemiology,
and End Results (SEER) Program data in the state of Louisiana among different
counties, which provide important information to study spatial survival rates. 

The remainder of the paper is organized as follows. In Section~\ref{sec:method},
we develop a homogeneity pursuit of survival data in the piecewise constant
proportional hazard framework  with gwCRP prior. In
Section~\ref{sec:bayes_comp}, a collapsed Gibbs sampler algorithm and post MCMC
inference are discussed. The extensive simulation studies are carried out in
Section~\ref{sec:simu}. For illustration, our proposed methodology is applied to
respiratory cancer survival data in Section~\ref{sec:real_data}. Finally, we
conclude this paper with a brief discussion in Section~\ref{sec:discussion}.

\section{Methodology}\label{sec:method}
\subsection{Spatial Piecewise Constant Hazards Models}

Let $T_{\ell i}$ denote the survival time for patient $\ell$ at location
$s_i$, with $\delta_{\ell i}=1$ representing the event and $\delta_{\ell i}=0$
indicating
censored,
and $X_\ell(s_i)$ denotes the vector of covariates corresponding to~$T_{\ell
i}$
for $i=1,2,...,n$, and $\ell=1,2,...,n_i$, where $n_i$ denotes the number 
of the patients at location $s_i$. In this paper, $s_1,s_2,\ldots,s_n$
are
areal
units which is defined in \citet{banerjee2014hierarchical}.
Let
$\bm{D}=\{(T_{\ell i},\delta_{\ell i},X_\ell(s_i)),
i=1,2,...,n,\ell=1,2,...,n_i\}$.
denote the observed data.
We consider a proportional hazards model \citep{cox1972} with
piecewise constant baseline hazard.
We partition $[0, \infty)$ into $J$ intervals ($0 = a_0 < a_1 < \dots
<
a_{J} = \infty$),
then the hazard function is given by
\begin{equation}\label{eq:hazard}
\lambda(t|X_{\ell}(s_i))=\lambda_{0}(t) \exp
(X_\ell(s_i)^\top
\bbeta), \quad
\end{equation}
with piecewise constant baseline hazard function
$\lambda_{0}(t)=\lambda_{j}$ for $a_{j-1}\leq t< a_{j}, 
~j=1,\ldots, J$.

For the piecewise constant hazard function mentioned in \eqref{eq:hazard}, the
baseline hazards $\lambda_1, \dots, \lambda_{J}$ and regression coefficients
$\bbeta$ are constants over different regions. Due to observed environmental
factors, spatially varying patterns in baseline hazards and regression
coefficients of hazard function need to be considered. The piecewise constant
hazard function with spatially varying pattern is therefore given by
\begin{equation}
\lambda(t|X_{\ell}(s_i))=\lambda_{0(s_i)}(t) \exp
(X_\ell(s_i)^\top \bbeta(s_i)), \quad
\label{eq:spatial_hazard}
\end{equation}
where $\lambda_{0(s_i)}(t)=\lambda_j(s_i)$ for $a_{j-1}\leq t< a_{j},
~j=1,\ldots,J$. Under this model, 
$\blambda(s_i)=(\lambda_1(s_i), \dots, \lambda_{J}(s_i))^{\top}$ and
$\bbeta(s_i)$ represent the location-specific baseline hazards and regression
coefficients.

After some algebra, the logarithm of likelihood function for observed survival
data $\bm{D}$ is obtained as
\begin{equation}\label{loglik}
\begin{split}
&\log\mathcal{L}(\bbeta(s_i), \blambda(s_i), i=1,\ldots,n\mid\bm{D})\\
&=\sum_{i=1}^n \left\{\sum_{j=1}^Jd_{ji} \log \lambda_j(s_i) +
\sum_{\ell=1}^{n_i} \delta_{\ell i} X_{\ell}(s_i)^\top\bbeta(s_i)
- \sum_{j=1}^J \lambda_j(s_i) \left[\sum_{\ell=1}^{n_i}
\Delta_{j}(T_{\ell i})\exp(X_{\ell}(s_i)^\top\bbeta(s_i))\right]\right\},	
\end{split}
\end{equation}
where
$
d_{ji}=\sum_{\ell=1}^{n_i} \delta_{\ell i} \1_{[a_{j-1}, a_j)}(T_{\ell i})
$,
which represents the number of people at location $s_i$ who
experience the event during the time period from $a_{j-1}$ to $a_j$,
and
$\Delta_{j}(t)=
t-a_{j-1} \text{ for } a_{j-1} \le t < a_j.
$

For one particular location $s_i$, let 
$\bm{\eta}(s_i)=\log \blambda(s_i)$ and define $\btheta(s_i)=(\bbeta(s_i)^\top,
\bm{\eta}(s_i)^\top)^\top$ the
collection of parameters, then the maximized likelihood estimate (MLE)
$\hat{\btheta}(s_i)$ can be obtained by solving the score function, which is
the derivative of the
logarithm
of likelihood function
in (\ref{loglik}), and the estimated variance-covariance matrix of MLE is 
$\hat{\Sigma}_i=(-H)^{-1}$, respectively, where~$(-H)$
denotes
the
negative Hessian matrix. Based on the MLEs and estimated variance-covariance
matrices, we have the following approximation of the likelihood.

\begin{proposition}\label{prop:normal_approximate}
We assume the regularity conditions A-D in \citet{friedman1982piecewise}.
As $n_i\rightarrow \infty,\, i=1\ldots,n$, the data likelihood
$\mathcal{L}(\bbeta(s_i), \blambda(s_i), i=1,\ldots,n\mid\bm{D})$ is
approximated as
	\begin{equation}
\mathcal{L}(\bbeta(s_i), \blambda(s_i), i=1,\ldots,n\mid\bm{D})\approx
\prod_{i=1}^n \text{MVN}(\hat{\btheta}(s_i)|\btheta(s_i),\hat{\Sigma}_i),
	\end{equation}
	where MVN stands for the multivariate normal distribution.
	\end{proposition}

The derivations of $\hat{\Sigma}_i$ and the proof of
Proposition \ref{prop:normal_approximate} are given in 
Section A and Section B of Supporting information.
Instead of using the log likelihood in
\eqref{loglik}, our following model is based on normal approximation in
Proposition
\ref{prop:normal_approximate} for computational convenience.

Based on the normal approximation given in Proposition
\ref{prop:normal_approximate}, a natural way which follows
\citet{gelfand2003spatial} for spatially varying pattern of baseline hazards
and regression is to give a Gaussian process prior to ${\btheta}(s_i),\,
i=1,\ldots,n$. The Gaussian process for ${\btheta}(s_i),\, i=1,\ldots,n$ is
defined as
\begin{equation}
\btheta\sim \text{MVN}(\bm{1}_{n\times1}\otimes\bm{\mu},\bm{H}(\phi)\otimes
\Sigma),
\end{equation}
where $\btheta=(\btheta(s_1)^\top,\ldots,\btheta(s_n)^\top)^\top$, $\bm{\mu}$
is a $p+J$ dimensional vector, $\bm{H}(\phi)$ is a $n\times n$ spatial
correlation matrix depending on the distance matrix with parameter $\phi$,
$\Sigma$
is a $(p+J) \times (p+J)$ covariance matrix, and $\otimes $ denotes Kronecker
product. The $(i,j)$-th entry of $\bm{H}(\phi)$ is $\exp(-\phi|s_i-s_j|)$,
where
$|s_i-s_j|$ is the distance between $s_i$  and $s_j$, and
$\phi>0$ is the range
parameter for spatial correlation. For the Gaussian process prior, the
parameters of closer locations have stronger correlations.

For many spatial survival data, some regions will share same covariate effects
or baseline hazards with their nearby regions. In addition, some regions will
share similar parameters regardless of their geographical distances, due to the
similarities of regions' demographical information such as income distribution
\citep{ma2019bayesian,hu2020bayesian}, food environment index, air pollution
\citep{zhao2020bayesian}, and etc.. A spatially varying pattern
for ${\btheta}(s_i),\, i=1,\ldots,n$ is not always valid. Based on
the homogeneity pattern, we focus on the clustering of spatially-varying
parameters. In our setting, we assume that the $n$ parameter vectors can be
clustered into $k$ groups, i.e., $\btheta(s_i)=\btheta_{z_i}$ where $z_i \in
\{1,2,\ldots,k\}$.

\subsection{Geographically Weighted Chinese Restaurant Process}
A latent clustering structure can be introduced to accommodate the
spatial heterogeneity on parameters of sub-areas.
Under the frequentist framework, the clustering problem 
could be solved in a two-stage approach: first obtain the estimate of 
number of clusters, $\hat{k}$,
then detect the optimal clustering assignment 
among all possible clusterings of $n$ elements into $\hat{k}$ clusters.
However, in this approach, the performance of the estimation of 
cluster assignments highly relies
on the estimated number of clusters, it may ignore
uncertainty in the first stage and cause redundant cluster assignments.
Bayesian nonparametric method is a natural remedy to simultaneously estimate
the number of clusters and cluster assignments. The Chinese restaurant 
process \citep[CRP;][]{pitman1995exchangeable, neal2000markov}
offers choices to allow uncertainty in the number of clusters by assigning a
prior distribution on $(z_1, z_2, \ldots, z_n)$.
In CRP, $z_i,~i=2, \ldots, n$ are defined through the
following
conditional
distribution
\citep[also called a P\'{o}lya urn scheme,][]{blackwell1973ferguson}.
\begin{eqnarray}\label{eq:crp}
P(z_{i} = c \mid z_{1}, \ldots, z_{i-1})  \propto   
\begin{cases}
\abs{c}  , &\text{at an existing cluster labeled}\, c,\\
\alpha,  &\text{at a new cluster}.  
\end{cases}
\end{eqnarray}
Here $\abs{c}$ refers to the size of cluster labeled $c$, and $\alpha$ is the
concentration parameter of the underlying
Dirichlet process. 
Based on the P\'{o}lya urn scheme shown in \eqref{eq:crp}, the
customers will have no preference for sitting with different customers. For
the spatial survival data, nearby regions will share similar environmental
effects such as P.M. 2.5, water quality, etc.. These similar effects will lead
the nearby sub-regions to share similar parameters. In order to consider
similar effects caused by geographical distance,
we modify the traditional CRP to
geographically weighted CRP (gwCRP) so that the customer will have
higher probability sitting with their familiar customers which are
geographically nearby.
We have the conditional distribution of $\bm{\theta}(s_i)$ given
$\btheta(s_1),\ldots,\btheta(s_{i-1})$ based on following definition.
\begin{definition}\label{def:conditional}
If $G_0$ is a continuous distribution and $i>1$, the distribution of
$\bm{\theta}(s_i)$ given $\bm{\theta}(s_1),\ldots,\bm{\theta}(s_{i-1})$ is
proportional to
\begin{align}
	f(\bm{\theta}(s_i)\mid
	\btheta(s_1),\ldots,\btheta(s_{i-1})) 
	 \propto \sum_{r=1}^{K^*}
\sum_{j=1}^{i-1}
w_{ij}\bm{1}(\bm{\theta}(s_j)=\bm{\theta}^*_r)\delta_{\bm{\theta
}^*_r}
(\bm{\theta}(s_i)) + \alpha G_0(\bm{\theta}(s_i)),
\label{eq:conditional_distribution}
\end{align}
where 
$f(\cdot)$ is the distribution density function,
$K^*$ denote the number of clusters excluding the $i$-th observation,
$\bm{\theta}^*_1,\ldots,\bm{\theta}^*_{K^*}$ are $K^*$ distinguished values of
$\bm{\theta}_1,\ldots,\bm{\theta}_{i-1}$, $w_{ij}$ is geographical weight
which is calculated by the
distance between $s_i$ and $s_j$, and $\delta(\cdot)$ is the
Dirac measure.
\end{definition}

Based on the Definition \ref{def:conditional}, we have similar P\'{o}lya urn
scheme called gwCRP for conditional distribution in
\eqref{eq:conditional_distribution} with CRP.

\begin{proposition}\label{prop:gwcrp}
	A P\'{o}lya urn scheme of gwCRP is defined as
	\begin{eqnarray}\label{eq:gwcrp}
P(z_{i} = c \mid z_{1}, \ldots, z_{i-1})  \propto   
\begin{cases}
\abs{c^*}  , &  \text{at an existing cluster labeled}\, c,\\
\alpha,  & \text{at a new cluster}.  
\end{cases}
\end{eqnarray}
$\abs{c^*}=\sum_{j=1}^{i-1}w_{ij}\1(z_j=c)$, where $w_{ij}$ is the geographical
weight.
\end{proposition}

Compared the existing geographically weighted regression
literatures, our weights are obtained by graph distance between different
areas. Following
\cite{xue2019geographically},
we
denote a graph as $G$, with set of vertices $V(G)=\{v_1,\ldots, v_n\}$, and set
of edges $E(G) = \{e_1,\ldots, e_m\}$. The graph distance between two vertices
$v_i$ and $v_j$ is defined as:
\begin{equation}
d_{v_i v_j}=\begin{cases}
|V(e)|, &\text{if } e \text{ is the shortest path connecting } v_i
\text{ and } v_j, \\
\infty, &\text{if } v_i \text{ and }\, v_j \text{ are not connected},
\end{cases}
\label{eq:netdistance}
\end{equation}
where $|V(e)|$ represents the cardinality of edges in $e$. For the county
level data, we construct the graph $G$ based on adjacency matrix among
different counties. We treat $n$ counties as $n$ vertices of this graph and
$v_i$ and $v_j$ are connected when the corresponding counties share the
boundary. Based on the
graph distance calculated by \eqref{eq:netdistance}, we calculate the
geographical weights by:
\begin{equation}
w_{ij}=\begin{cases} 1,& \text{ if } d_{v_i v_j} \leq 1,\\
    \exp(-d_{v_i v_j}\times h), & \text{ if } 1 < d_{v_i v_j},
\end{cases}
\label{eq:stochastic_neighbor_weights}
\end{equation}
where $d_{v_iv_j}$ is the graph distance between areas $i$ and $j$. For the
weighting function in
\eqref{eq:stochastic_neighbor_weights}, we give the largest weight
($w_{ij}\equiv 1$) for the areas sharing the same boundaries,
which
follows the first law of geography~\citep{tobler1970computer}. 
For simplicity, we refer to gwCRP introduced above as
$\text{gwCRP}(\alpha,h)$, where $\alpha$ is the concentration parameter for
Dirichlet distribution and $h$ is the spatial smoothness parameter.
\begin{remark}
\label{remark:1}
Based on the P\'{o}lya urn scheme defined in \eqref{eq:gwcrp} and geographical
weighting scheme defined in \eqref{eq:stochastic_neighbor_weights}, we find
that (i) when $h=0$, the gwCRP reduces to traditional CRP, which
leads to over-clustering problem in estimating of the number of clusters;
(ii)
when~$h \rightarrow \infty$, a new customer just only choose the table
representing
spatially contiguous regions. This will also lead to the same over-clustering
problem as CRP.
\end{remark}

\subsection{gwCRP for Piecewise Constant Hazards Models}

Adapting gwCRP to the piecewise constant hazards models, our model and prior
can be expressed hierarchically as:
\begin{align}\label{eq:MFMNPP}
\begin{split}
&\log\mathcal{L}(\bbeta_{z_i}, \blambda_{z_i}, i=1,\ldots,n\mid\bm{D})\\
&=\sum_{i=1}^n \left\{\sum_{j=1}^Jd_{ji} \log \lambda_{jz_i} +
\sum_{\ell=1}^{n_i} \delta_{\ell i} X_{\ell}(s_i)^\top\bbeta_{z_i}
- \sum_{j=1}^J \lambda_{jz_i} \left[\sum_{\ell=1}^{n_i}
\Delta_{j}(T_{\ell i})\exp(X_{\ell}(s_i)^\top\bbeta_{z_i})\right]\right\},\\
& z_i\mid \bm{\pi},k \sim \text{Multinomial}(\pi_1,\cdots,\pi_k),  \\
& \bm{\pi}\sim \text{gwCRP}(\alpha,h), \\
& \bm{\theta}_r
\sim \mbox{MVN}(0, \Sigma_0),  \quad r= 1, \ldots, k,
\end{split}
\end{align}
where
$\btheta_r=(\beta_{1r},\ldots,\beta_{pr},\log \lambda_{1r},\ldots,\log
\lambda_{Jr})^\top$ is a $p+J$ dimensional vector. And let $k\rightarrow
\infty$, and $\Sigma_0$ be hyperparameter
for base distribution of $\btheta_1,\ldots,\btheta_r$. We choose
$\Sigma_0=100\bm{I}$ in all the simulations and real data analysis providing 
noninformative priors. 
The concentration parameter $\alpha$ controls the 
probability of introducing a new cluster which is similar with CRP. Different 
values of $h$ lead to different weighting scale for different sub-regions. In 
our following simulations and real data analysis, we fix $\alpha=1$ and tune 
$h$ with different values.

\section{Bayesian Inference}\label{sec:bayes_comp}

In this section, we will introduce the MCMC sampling algorithm, post MCMC
inference method, and Bayesian model selection criterion.

Our goal is to sample from the posterior distribution of the unknown parameters
$k$, $\bm{z} = (z_1,...,z_n) \in \{1,...,k\}^n$,
$\bm{\beta}=(\bm{\beta}_1,\ldots,\bm{\beta}_k)$, and
$\bm{\lambda}=(\bm{\lambda}_1,\ldots,\bm{\lambda}_k)$.
Based on Proposition \ref{prop:normal_approximate} and Proposition
\ref{prop:gwcrp}, we can efficiently cycles through the full
conditional distributions of $z_i | z_{-i}$ for $i = 1,2,\ldots,n$ and
$\bm{\beta}^\top,\log \bm{\lambda}^\top$, where $z_{-i} =
\bm{z}\setminus{\{z_i}\}$.
The
marginalization over $k$ can avoid
complicated reversible jump MCMC algorithms or even allocation samplers. The
full conditionals of $z_1,\ldots,z_n$ are given in Proposition
\ref{prop:full_conditional}. The details of sampling algorithm is given in
Section C of Supporting information.

\begin{proposition}\label{prop:full_conditional}
The full conditional distributions $P(z_i = c \mid z_{-i}, \hbtheta, \btheta)$
of $z_1,\ldots,z_n$ is given as
\[\propto \left\{
\begin{array}{ll}
& \left(\sum_{j\neq i}w_{ij}\1(z_j=c)\right)
(2\pi)^{-\frac{p}{2}}|\hat{\Sigma}_i|^{-\frac{1}{2}}\exp\left\{-\frac{1}{2
}\left(
(\hbtheta(s_i)-\btheta_c)^\top
\hat{\Sigma}_i^{-1}(\hbtheta(s_i)-\btheta_c)\right)\right\}
\text{   at existing $c$} \\
& \alpha
(2\pi)^{-\frac{p}{2}}|\hat{\Sigma}_i|^{-\frac{1}{2}}|\Sigma_0|^{-\frac{1}{2}}
|\hat{\Sigma}_i^
{-1}+\Sigma_0^{-1}|^{-\frac{1}{2}}\exp\left\{-\frac{1}{2}\left(\hbtheta(s_i)
^\top
(\hat{\Sigma}_i+\Sigma_0)^{-1} \hbtheta(s_i)\right)\right\} \text{  if $c$ is a
new
cluster} \\
\end{array} 
\right. \]
where $\hat{\Sigma}_i$ is the estimated variance-covariance matrix of MLE
$\hbtheta(s_i)$ for $i=1,\ldots,n$, and 
$\Sigma_0$ is the variance hyperparameter for the base
distribution of $\btheta_1,\ldots,\btheta_r$.
\end{proposition}

We carry out posterior inference on the group memberships $z_1,\ldots,z_n$ by
using Dahl's method
\citep{dahl2006model}, which proceeds as follows

\begin{enumerate}

\item {Define membership matrices $B^{(l)} =(B(i,j))_{i,j \in
\left\{1,\ldots,n\right\} } =  (\bm{1}(z_{i}^{(l)} = z_{j}^{(l)}))_{n \times
n}$, where $l = 1, \ldots, B$ indexes the number of retained MCMC draws after
burn-in, and $\bm{1}(\cdot)$ is the indicator function.}
\item {Calculate the average membership matrix $\bar{B} = \frac{1}{B}
\sum_{l=1}^{L}B^{(l)}$, where the summation is element-wise.}
\item {Identify the most \emph{representative} posterior sample as the one
that
is closest to $\bar{B}$ with respect to the element-wise Euclidean distance
$\sum_{i=1}^{n} \sum_{j=1}^{n} (B^{(l)}(i,j) - \bar{B}(i,j))^{2}$ among the
retained $l = 1,\ldots,B$ posterior samples.}
\end{enumerate}

Therefore, the posterior estimates of cluster memberships $z_1,\ldots,z_n$ and
model parameters~$\bm{\theta}$ can be obtained based on the draw identified by
Dahl's
method.

We recast the choice of decaying parameter $h$ as a model selection problem. We
use the Logarithm of the Pseudo-Marginal
Likelihood \citep[LPML;][]{ibrahim2001bayesian} based on
conditional predictive ordinate
\citep[CPO;][]{gelfand92modeldetermination,seisser1993,gelfand1994bayesian} to
select
$h$. The LPML is
defined as
\begin{equation}
	\text{LPML} = \sum_{i=1}^{N} \text{log}(\text{CPO}_i),
	\label{eq:lpml}
\end{equation}
where $\text{CPO}_i$ is the i-th conditional predictive ordinate. 
Following \cite{chen2000}, a Monte
Carlo estimate of the CPO, within the Bayesian framework, 
can be obtained as
\begin{equation}
	\label{eq:CPOest}
\widehat{\text{CPO}}_i^{-1} = \frac{1}{B} \sum_{b=1}^{B}
\frac{1}{f(D_i|\btheta_{z_i}^b)},
\end{equation}
where $B$ is the total number of Monte Carlo iterations, $\btheta_{z_i}^b$ is
the $b$-th posterior sample, 
and $f(\cdot)$ is the
likelihood function define in~\eqref{loglik}. An estimate of the
LPML can subsequently be calculated as:
\begin{equation}
	\label{eq:LPML}
	\widehat{\text{LPML}} = \sum_{i=1}^{N} \text{log}(\widehat{\text{CPO}}_i).
\end{equation}
A model with a larger LPML value is preferred.

\section{Simulation}\label{sec:simu}
\subsection{Simulation Setting and Evaluation Metrics}

\begin{figure}
    \centering
    \includegraphics[width=1\textwidth]{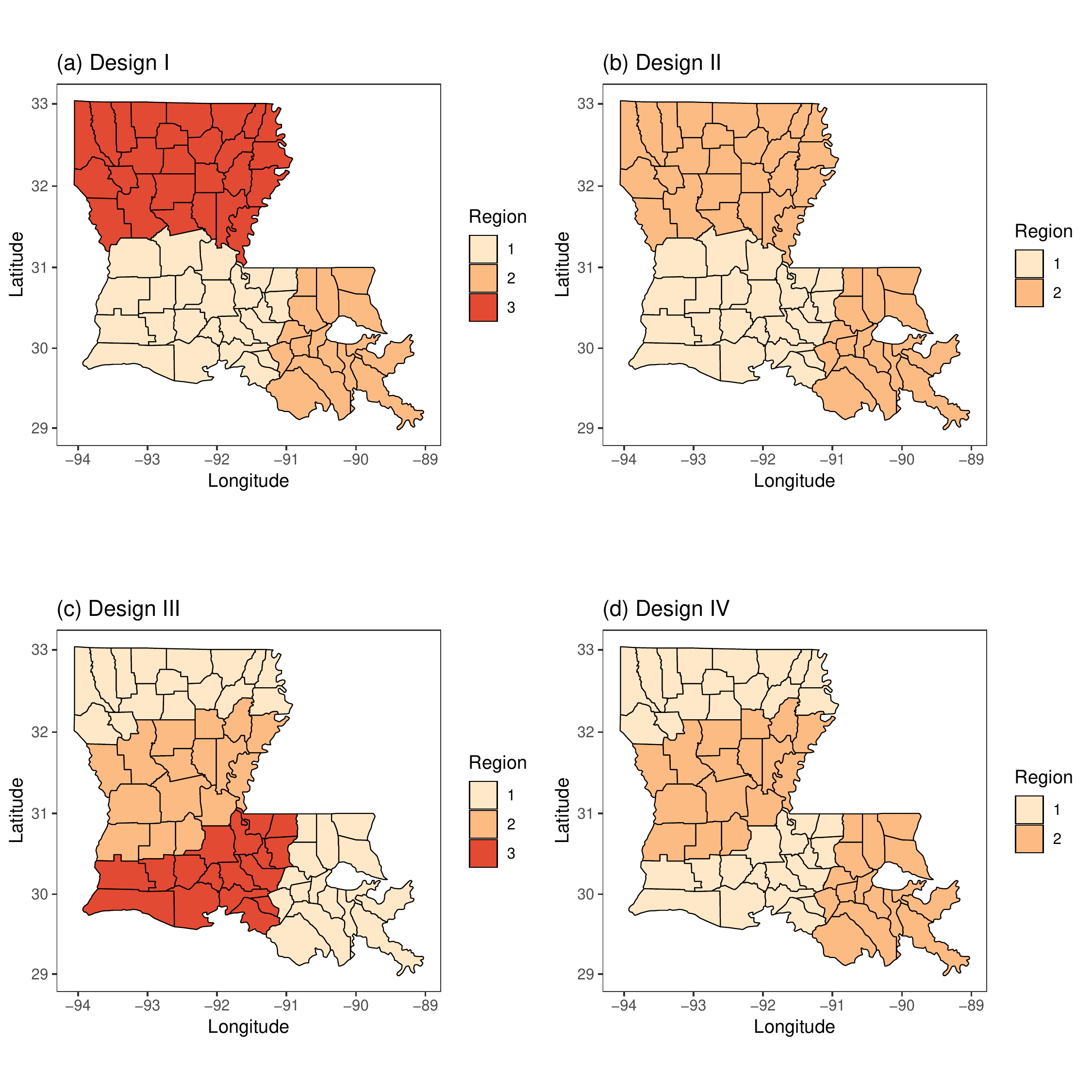}
    \caption{Geographical clustering patterns in 
Louisiana state of simulation designs (This figure appears in color in the
electronic version of this article, and any mention of color refers to that
version.)}
    \label{fig:simu_design}
\end{figure}

In this section, we present simulation studies under four different designs
to illustrate the performance of
our proposed gwCRP method and compare with traditional CRP, 
in terms of both clustering configuration 
and estimation of regression coefficients and piecewise constant baseline
hazards under proportional hazards
model. Survival datasets that resemble the SEER respiratory cancer data for
Louisiana are generated. The censoring rate is around~30\%.
We design four different geographical clustering patterns
in Louisiana state, which are shown in Figure~\ref{fig:simu_design}. Designs I
and III have three true clusters, and Designs II
and IV have two true clusters. 
In addition, Designs II and III both have one cluster
consisting of two disjoint areas since, in practice,
it is still possible for two
distant counties to belong to the same cluster.
Design IV has two clusters both consisting of disjoint areas.

For each design, 100 replicate datasets are generated
under proportional hazards model with piecewise constant baseline hazard.
In each replicate, we generate survival data of 60 subjects for each county,
including three regression covariates from $N(0,1)$ i.i.d.,
survival time and censoring. 
We set three pieces for the baseline hazards with cutting points 1.5 and 6 for
all designs,
and over four designs, we
have three true clusters at maximum, 
and the true regression coefficients and baseline hazards  
used are chosen from
$\bbeta_1 = (1, 0.5, 1), \blambda_1 = (0.045,0.036,0.045)$,
$\bbeta_2 = (1.5, 1, 1), \blambda_2 = (0.045,0.036,0.036)$,
and $\bbeta_3 = (2, 0.5, 1.5), \blambda_3 = (0.036,0.045,0.0495)$. 
Censoring times are generated independently by taking the minimum of 
150 and random values from Exp(0.01) with expectation 100.
For each replicate, we set $\alpha=1$ 
and run different values of $h$, from 0 to 2 with grid 0.2, 
and from 3 to 10 with grid 1, and select the optimal $h$ via LPML.
A total of 2000 MCMC iterations are run for each replicate, 
with the first 500 iterations as burn-in.

To compare the performance of clustering
of gwCRP under different values of $h$, both estimation
of the number of clusters and the matchability of clustering
configurations are reported. In our simulation, we use
mean Rand Index \citep{rand1971objective} which is
obtained by using R-package
\textbf{fossil} \citep{vavrek2011fossil} to measure the clustering performance.

In addition to clustering performance, we further evaluate the
estimation performance of covariates coefficients and baseline
hazards, which is assessed by average of bias (AB) and 
average of mean squared error (AMSE)
defined as follows.
Let $\bm{z}=(z_1,\ldots,z_n)$ 
be the true clustering label vector,~$\btheta_r(s_i)$ be the true parameter
value of cluster $r$,
$\kappa_r=\sum_{i=1}^{64}\1(z_i=r)$ be the number of 
counties in cluster $r$, $r=1,\ldots,k$, 
$\sum_{r=1}^k\kappa_r=n$, and for simulated data set $t$, 
let $\hat{\btheta}_{(t)}(s_i)$ be Dahl's
method estimate at location $s_i$ for $t$-th replicates.
Then AB is calculated as
\begin{equation*}
\text{AB}=\frac{1}{k}
\sum_{r=1}^k\frac{1}{\kappa_r}
\sum_{i|z_i=r}
\frac{1}{100}\sum_{t=1}^{100}
(\hat{\btheta}_{(t)}(s_i)-\btheta_r(s_i)),
\end{equation*}
and AMSE is calculated as
\begin{equation*}
\text{AMSE}=\frac{1}{k}
\sum_{r=1}^k\frac{1}{\kappa_r}
\sum_{i|z_i=r}
\frac{1}{100}\sum_{t=1}^{100}
(\hat{\btheta}_{(t)}(s_i)-\btheta_r(s_i))^2,
\end{equation*}
which calculates mean squared errors for each cluster first,
and then average across clusters.

\subsection{Simulation Results}

\begin{figure}[H]
    \centering
    \includegraphics[width=0.48\textwidth]{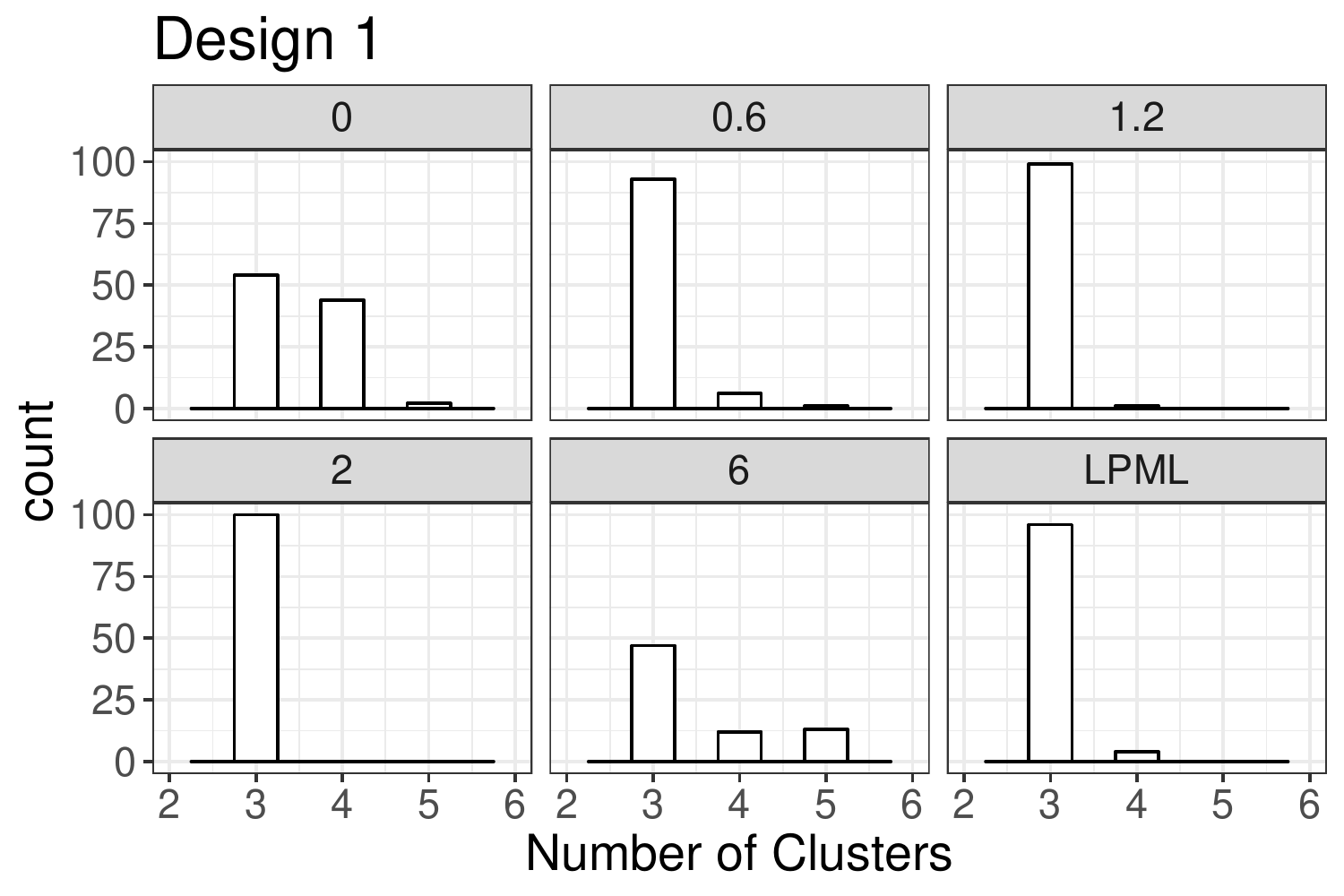}
    \includegraphics[width=0.48\textwidth]{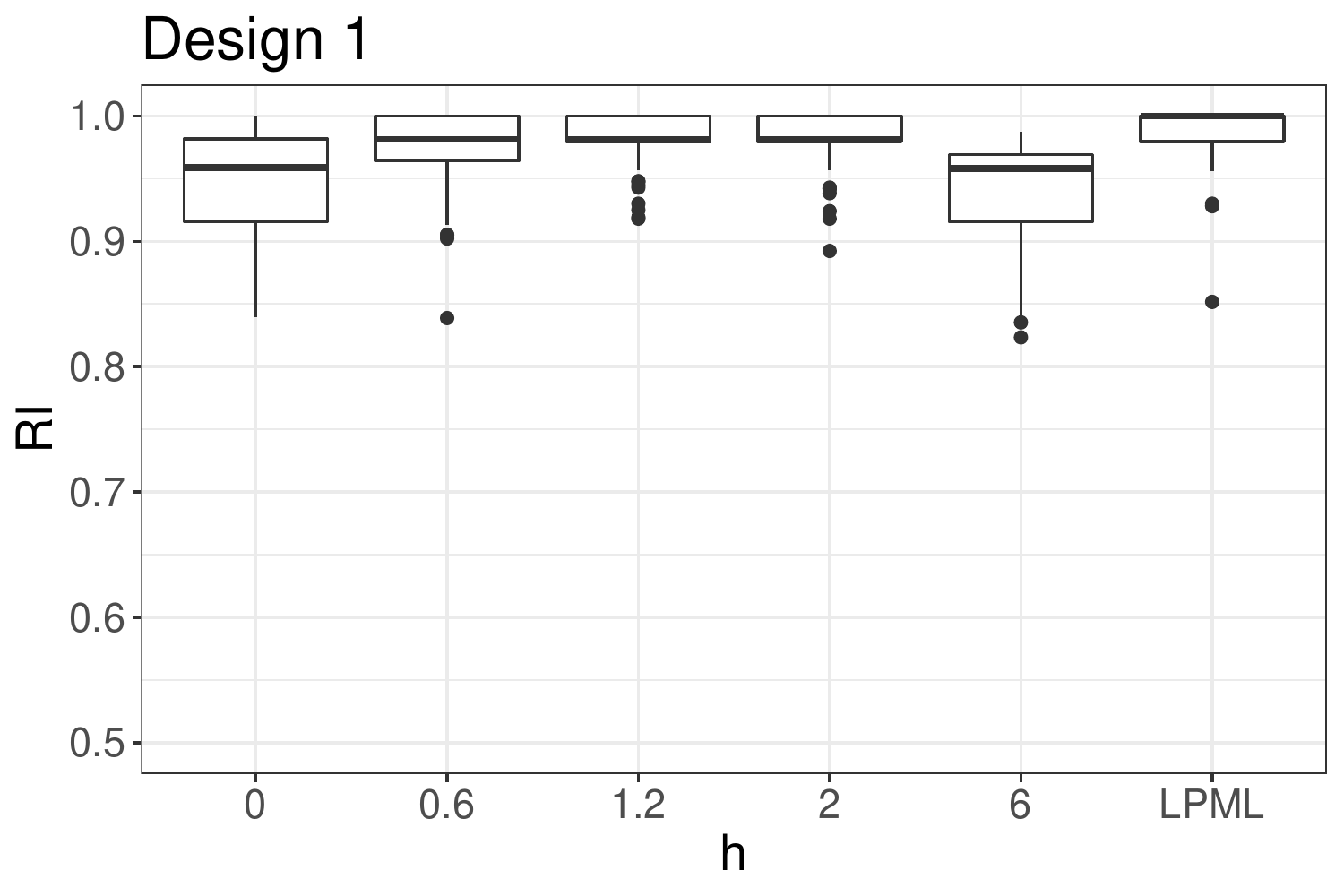}
    \includegraphics[width=0.48\textwidth]{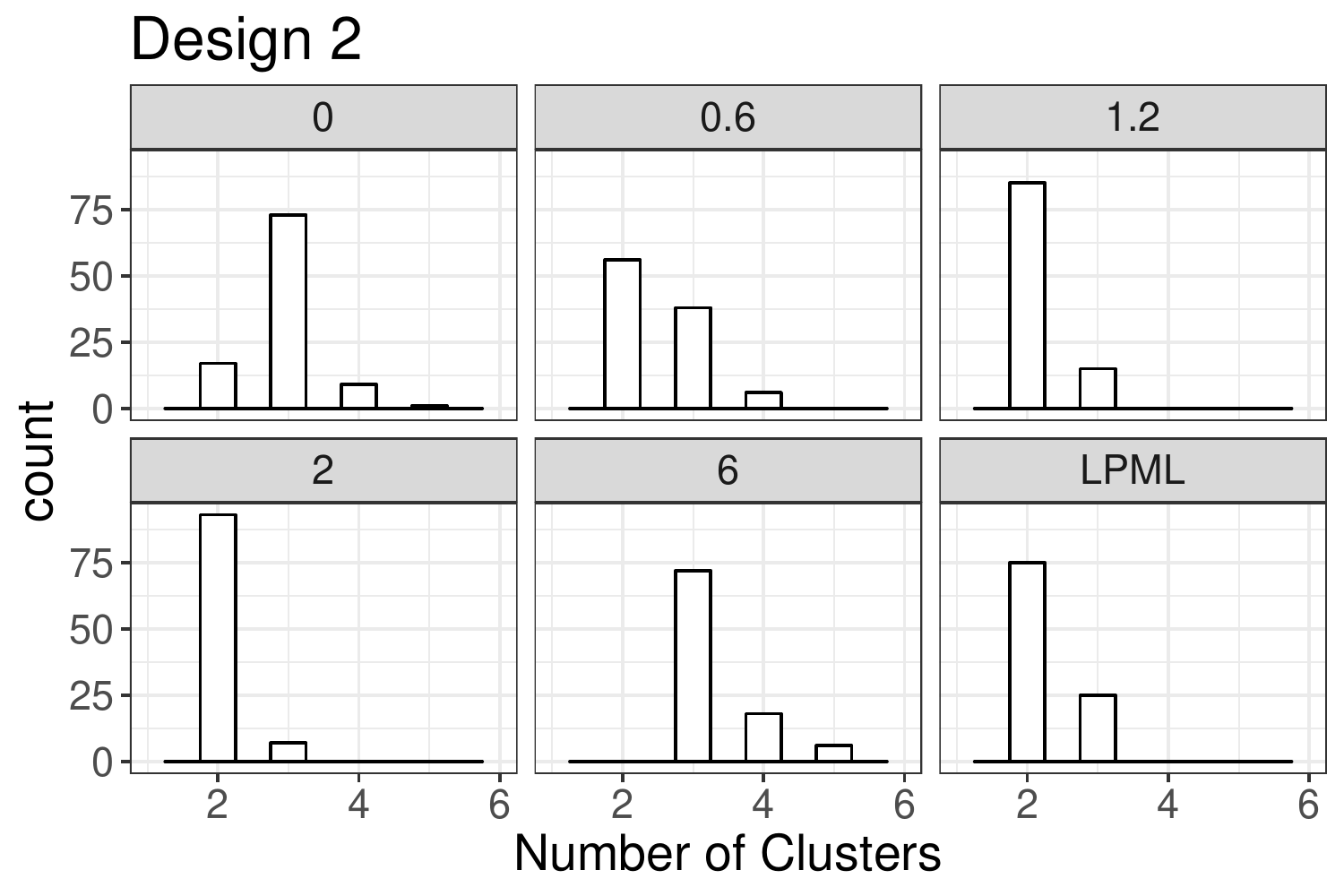}
    \includegraphics[width=0.48\textwidth]{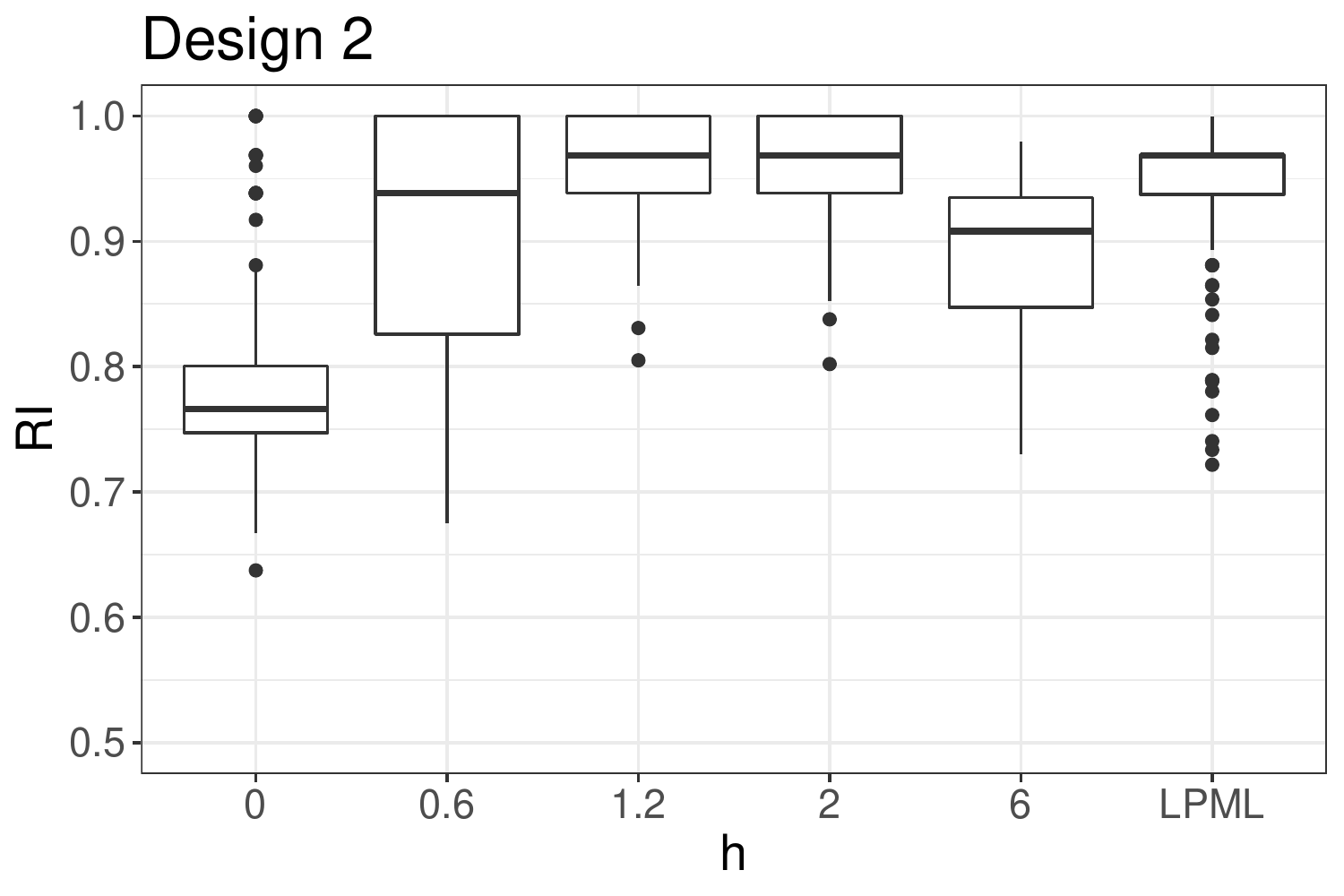}
    \includegraphics[width=0.48\textwidth]{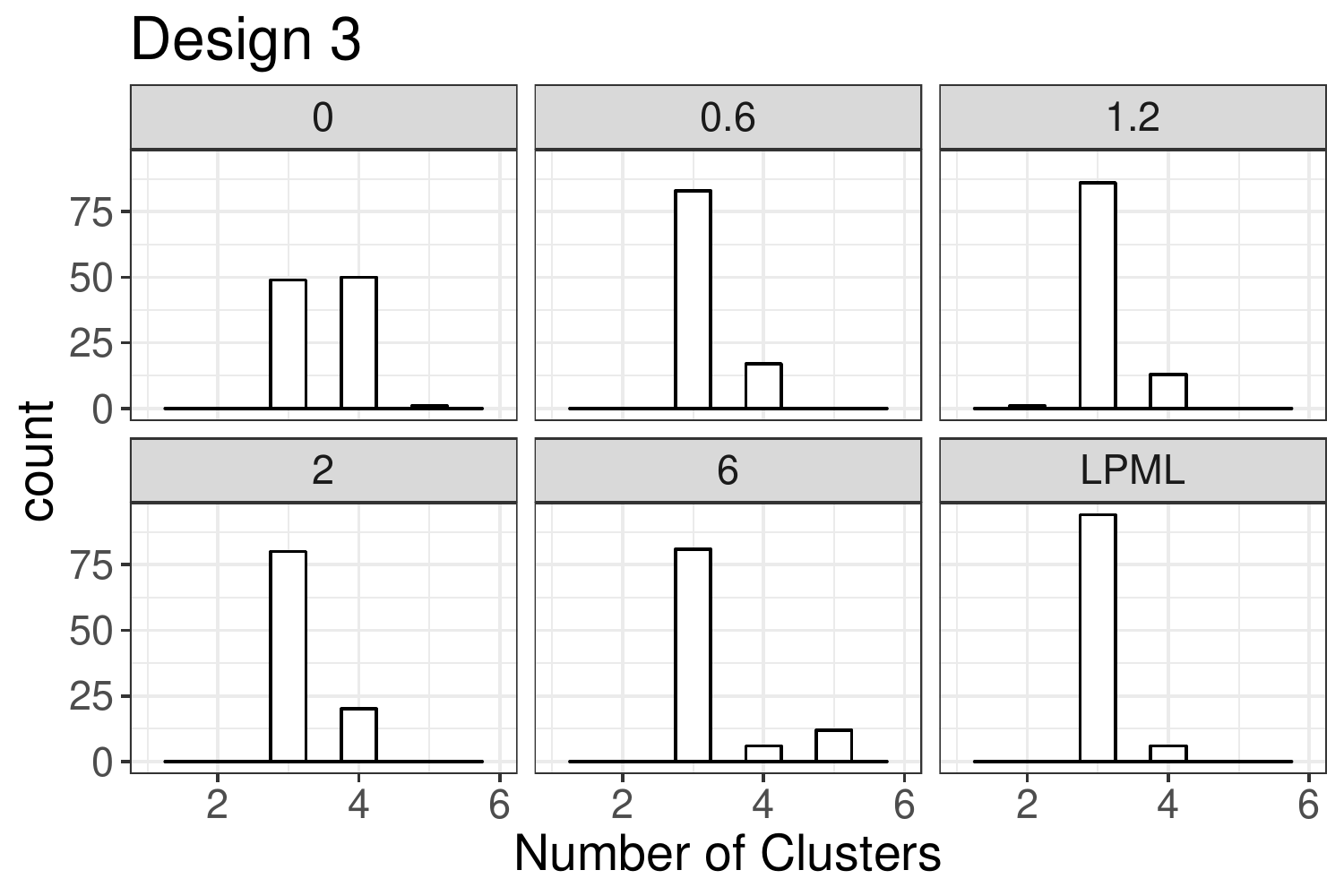}
    \includegraphics[width=0.48\textwidth]{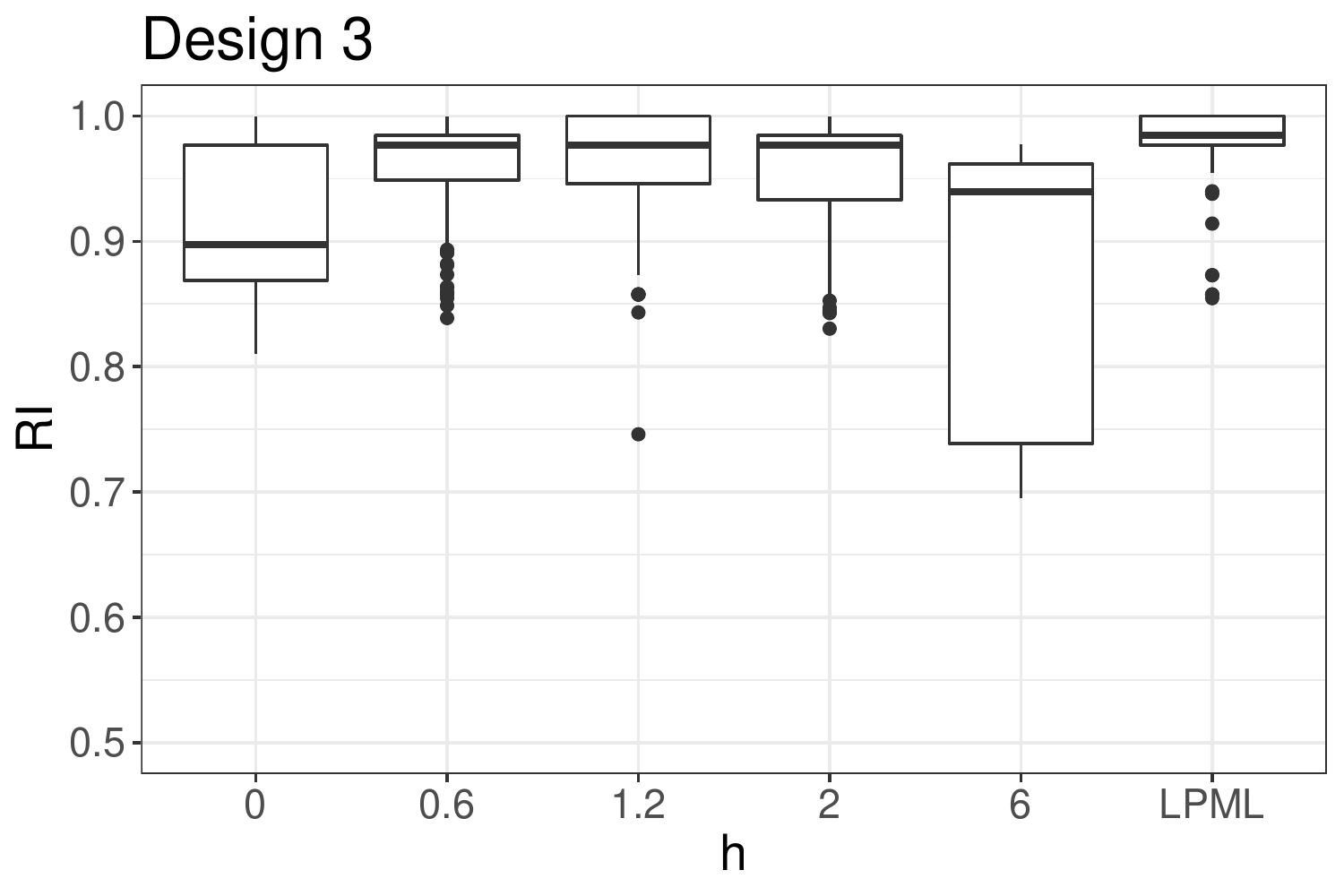}
    \includegraphics[width=0.48\textwidth]{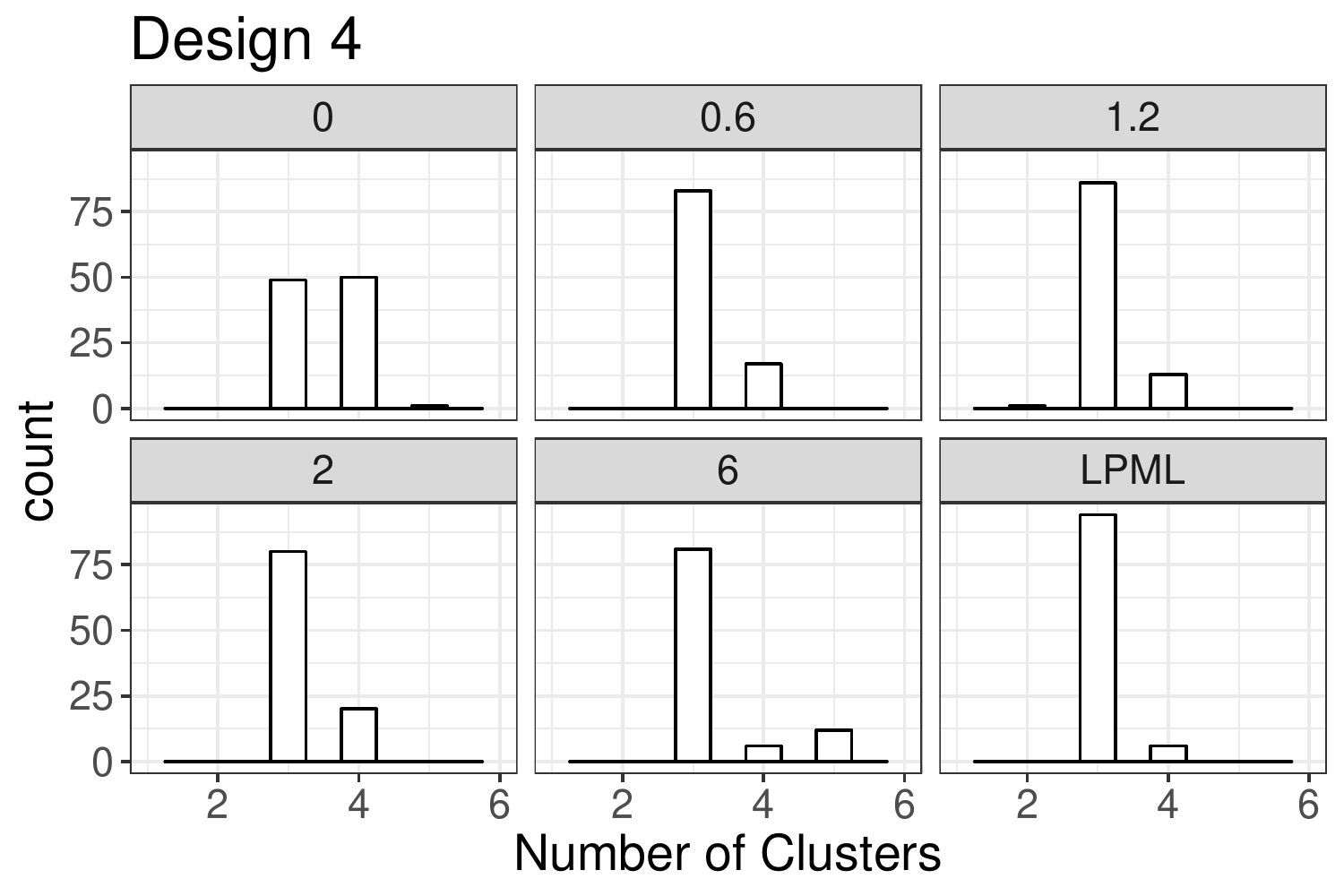}
    \includegraphics[width=0.48\textwidth]{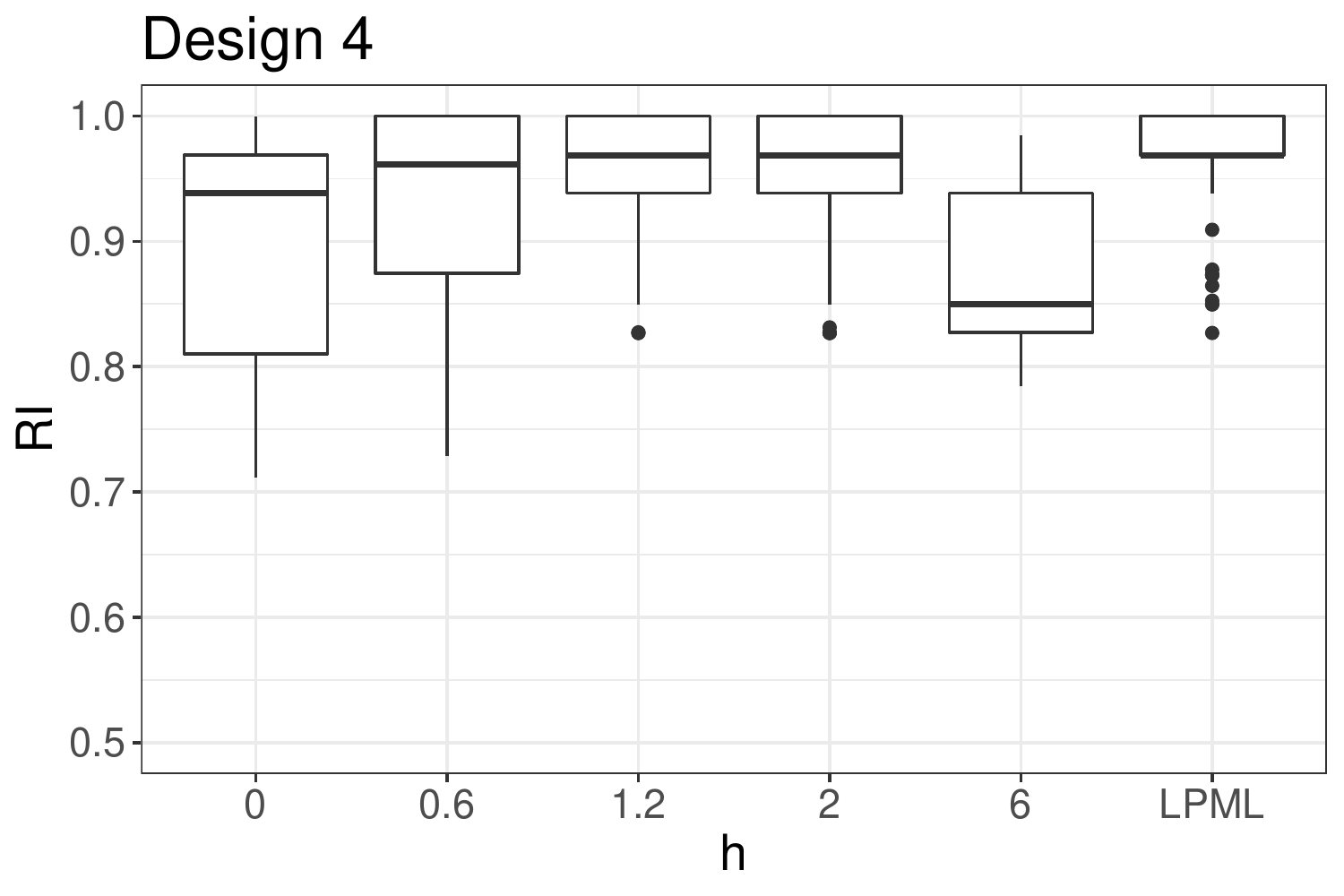}
    \caption{Histogram of estimates of $k$ and boxplot 
    of Rand Index under different $h$ and LPML selection 
    for simulation designs. 
    The average h selected by LPML is 1.296 in Design 1, 
    1.412 in Design 2, 1.366 in Design 3, 1.602 in Design 4.}
    \label{fig:simures_design1}
\end{figure}

Figure \ref{fig:simures_design1} shows the histogram of $k$ estimates
and boxplots of Rand Index
under different $h$ and the optimal selected by LPML for four simulation
designs. We see that when $h=0$, the proposed gwCRP method is 
identical to the traditional CRP method, and in this case, 
CRP always tends to over-cluster and often yields smaller Rand Index than the
results under~$h>0$. Another important trend is that, as $h$ increases, 
the estimated number of clusters decreases first and then increases, 
and the Rand Index increases first and then decreases as $h$ becomes too large.
As we discussed in Remark~\ref{remark:1}, this is because 
when $h$ increases from 0, the spatial patterns in
the data is captured by the proposed gwCRP method. 
However, as $h\rightarrow\infty$, 
the geographical weights $w_{ij}$ for spatial-discontiguous 
counties become 0,
which means only adjacent counties can be classified into the same cluster,
therefore leading to over-clustering phenomenon again. 
It is also discovered that the clustering perfomance under optimal $h$
selected by LPML is very well, with the probability of selecting true number 
of clusters always greater than 0.75, 
and Rand Index larger than or similar to the highest results attained by some
fixed value of $h$.

Table~\ref{tab:simures_AMSE_AB} summarizes the AB and AMSE results of
estimating
parameters of gwCRP under different $h$ for different designs.
For simplicity of summary results, here the AMSE of $\bbeta$ is the average of
AMSE of $\beta_1,\beta_2,\beta_3$ since they have similar scales, and the value
of $\log\blambda$ is the
average of AMSE of 
$\log\lambda_1,\log\lambda_2,\log\lambda_3$, respectively.

For Designs II, III and IV, the absolute value of AB 
decrease as $h$ increase from 0 to moderate values,
and increase again as $h$ increase to relatively large values. 
For all four designs, the absolute values of AB for 
$\blambda$'s of optimal $h$ selected by LPML 
always are the smallest, and the absolute values for 
both $\bbeta$'s and $\blambda$'s of
optimal $h$ selected by LPML are always smaller than the 
values of traditional CRP. 
The patterns in AMSE are more clear when comparing different
methods, that
traditional CRP has the largest AMSE
and AMSE decrease as $h$ increase from 0 to moderate values,
and increase again as $h$ increase to relatively large values. 
The results of optimal $h$ selected by LPML also has the best performance in
estimation.

\begin{sidewaystable}
	\centering
	\caption{AB and AMSE for parameter estimation under 
	different $h$ and LPML selection 
    for different simulation designs}
    \label{tab:simures_AMSE_AB}
	\begin{tabular}{lccccccccc}
		\toprule
		Method & Parameter & \multicolumn{2}{c}{Design I} & 
		\multicolumn{2}{c}{Design II} & 
		\multicolumn{2}{c}{Design III} &
		\multicolumn{2}{c}{Design IV} \\
\midrule
 & & AB & AMSE & AB & AMSE & AB & AMSE
 & AB & AMSE\\\cline{3-10}
\text{gwCRP $h=0.6$}  
& $\bbeta$  & -0.0063 & 0.0069 &0.0038 &0.0067 
&-0.0065 &0.0083 &-0.0027 &0.0060\\
&$\blambda$ &0.0815 &0.0193 &0.0732 &0.0186 
&0.0797 &0.0219 &0.0827 &0.0190\\ [1ex]
\text{gwCRP $h=1.2$}  
&$\bbeta$  &-0.0078 &0.0065 &-0.0002 &0.0058 
&-0.0061 &0.0087 &-0.0017 &0.0049\\
&$\blambda$ &0.0818 &0.0194 &0.0775 &0.0171 
&0.0794 &0.0217 &0.0815 &0.0172\\ [1ex]

\text{gwCRP $h=2$}  
&$\bbeta$  &-0.0096 &0.0068 &-0.0006 &0.0055 
&-0.0056 &0.0085 &-0.0051 &0.0049\\
&$\blambda$&0.0849 &0.0190 &0.0814 &0.0158 
&0.0798 &0.0216 &0.0873 &0.0191\\ [1ex]

\text{gwCRP $h=6$}  
&$\bbeta$  &-0.0061 &0.0129 &0.0030 &0.0072 
&-0.0299 &0.0204 &-0.0064 &0.0074\\
&$\blambda$ &0.0805 &0.0281 &0.0770 &0.0217 
&0.1005 &0.0296 &0.0852 &0.0195\\ [1ex]

\text{gwCRP Optimal}  
&$\bbeta$  &-0.0039 &0.0059 &0.0042 &0.0055 
&0.0074 &0.0067 &-0.0005 &0.0035\\
&$\blambda$ &0.0661 &0.0177 &0.0711 &0.0145 
&0.0732 &0.0203 &0.0777 &0.0177\\ [1ex]

CRP 
& $\bbeta$ &-0.0046 &0.0086 &0.0018 &0.0092 
&-0.0056 &0.0089 &0.0003 &0.0082 \\
&$\blambda$ &0.0760 &0.0228 &0.0717 &0.0233 
&0.0787 &0.0239 &0.0742 &0.0223\\ [1ex]

		\bottomrule
	\end{tabular}
\end{sidewaystable}

A sensitivity analysis regarding $\alpha$ and the weighting 
function is conducted. $\alpha=0.5, 1, 2, 5$ and the weighting function 
$w_{ij}=\exp(-d_{v_i v_j}^2\times h^2)\1\{d_{v_i v_j}>1\}+
\1\{d_{v_i v_j} \leq 1\}$
which has a faster decay to 0
are ran, and all the results are presented in Section D of Supporting
information.
The results show that the results of optimal $h$ selected by LPML are
insensitive to the choice of $\alpha$ and weighting function.

In a brief conclusion based on our simulation studies, gwCRP models have better
performance than CRP both for clustering and parameter estimation. Our proposed
model selection criterion, LPML, can nearly select the best performing $h$
value for both clustering and parameter estimation.

\section{SEER Respiratory Cancer Data}\label{sec:real_data}
\subsection{Data Description}

In this section, we apply our proposed model to analyze respiratory cancer data
in Louisiana state, which is downloaded from
the Surveillance, Epidemiology, and End Results (SEER) Program.We
analyzed the survival time of respiratory cancer patients using the SEER public
use data (SEER 1973-2016 Public-Use). We refer to \cite{mu2020bayesian}
for the detailed data clean description.
After cleaning, there
are 16213 observations left, and the censoring rate is 30.44\%. 
We select Age, Gender, Cancer grade and
Historical
stage of cancer for our analysis, and give the summary of 
survival times and covariates in Table \ref{tab:real_demo}.
The median survival times for patients in each county are provided in Section E
of Supporting information .

\begin{table}[tbp]
  \centering 
  \caption{Demographics for the studied dataset. For continuous variables, the
  mean and standard deviation (SD) are reported. For binary variables, the
  frequency and percentage of each class are reported.} \label{tab:real_demo}
  \begin{tabular}{lc}
    \toprule 
     & Mean(SD) / Frequency (Percentage) \\ \midrule 
     Survival Time & 22.43 (31.90) \\
     ~~Event & 12.63 (18.32) \\
     ~~Censor & 44.85 (43.06) \\
     Diagnostic Age & 66.55 (11.66)\\
     Sex \\
     ~~Female  & 6548 ($40.39\%$)\\
     ~~Male & 9665 ($59.61\%$)\\
     Cancer Grade \\
     ~~the class of lower grades  & 5307  ($32.73\%$)\\
    ~~the class of III or IV& 10906 ($67.27\%$)\\
     Historical Stage \tablefootnote{Distant stage means that a tumor has
spread to areas of the body distant or remote from the primary tumor} \\
     ~~not distant & 9005 ($55.54\%$)\\
     ~~distant & 7208 ($44.46\%$)\\
  \bottomrule 
  \end{tabular}
\end{table}

We first fit the Cox model of patients for each county using the covariates
selected. The regression coefficients are visualized in
Section E of Supporting information.
From results shown in Supporting information, it is seen that some counties
have similar
characteristics, no limited to only adjacent counties, indicating 
possibilities of globally discontiguous clusters.

\subsection{Data Analysis}

\begin{figure}[H]
    \centering
    \includegraphics[width=1.0\textwidth]{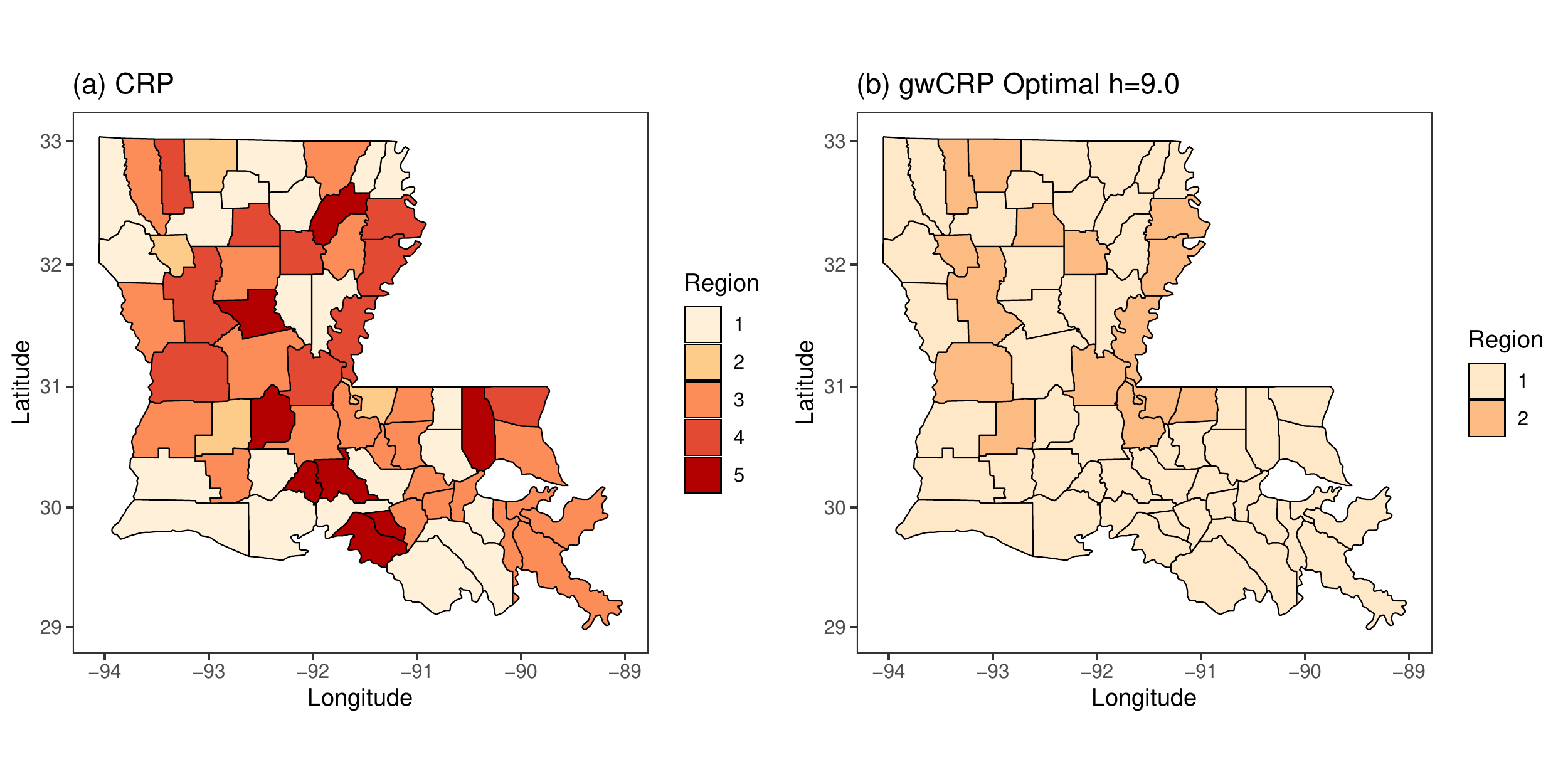}
    \caption{Clustering patterns of counties in Louisiana 
    state under CRP (when J=4) 
and gwCRP Optimal (when J=4, h=9.0) methods (This figure appears in color in
the electronic version of this article, and any mention of color refers to that
version.)}
    \label{fig:realdata_clustering}
\end{figure}

To select the optimal number of pieces for the baseline hazard and $h$, we run
$J=2$ with the cutpoints $(0, 9.01)$, 
$J=3$ with cutpoints $(0, 3.01, 9.01)$,
$J=4$ with cutpoints $(0, 1.01, 4.01, 9.01)$, 
and $J=5$ with cutpoints $(0, 1.01, 3.01, 5.01, 9.01)$. 
The cutpoints are set by dividing the start point $0$ and the median survival
time $9.01$ by quantiles evenly to ensure there are events at each piece for
each county.
For each $J$, we run $h$ from 0 to 10 with grid 0.1, and for each combination
of $J$ and $h$, 5000 MCMC iterations are run and drop the 
first 2000 as burn-in.
The optimal values selected by LPML is $J=4$ and $h=9.0$,
under which the corresponding estimate of
number of clusters is two, 
while the traditional CRP classifies the
counties into five clusters.
The trace plots of different chains of posterior samples of the estimates
for selected counties are presented in the Section E of
Supporting information to show the convergence of the MCMC.
The plots of clustering patterns
of CRP and gwCRP Optimal are shown in Figure
\ref{fig:realdata_clustering}, from which it is seen that
the gwCRP captures the globally discontiguous clusters very well.
The estimates and $95\%$ Credible Intervals of regression covariates
coefficients
and
baseline hazards obtained by gwCRP Optimal
are given in Table \ref{tab:real_est}, 
from which we see that,
though the baseline hazards are similar, the regression covariates coefficients
are quite different across different clusters. 
We see that our proposed method
successfully detects both spatially contiguous cluster and discontinuous
cluster simultaneously. The parameter estimates for Age are positive in all
counties, indicating that older patients on average are more likely to have the
event than younger patients. For the counties in cluster 1, the diagnostic ages
has higher hazards effects than other counties. However, for the counties in
Cluster 2, male, later cancer stage will have
higher hazards effects than other counties. The historical distance stage
effects are very similar in two clusters which indicates that the subjects with
tumor spreading will have higher hazards.

\begin{table}[H]
\centering 
  \caption{
  Estimates results of regression coefficients and
  baseline hazards obtained by gwCRP Optimal ($J=4,~h=9.0$).
  The $95\%$ Credible Interval for estimates of Cluster One is calculated by 
  the $95\%$ HPD Interval of County 13, and The $95\%$ Credible Interval for
estimates of Cluster Two is calculated by 
  the $95\%$ HPD Interval of County 33, where the counties were selected by 
  the minimum Euclidean distance from the posterior mean to the average
estimate.
  } 
  \label{tab:real_est}
\begin{tabular}{c|cc|cc}
\toprule
\multirow{2}{*}{Parameter} & 
\multicolumn{2}{c}{Cluster 1} & \multicolumn{2}{c}{Cluster 2}\\
\cline{2-5}
& Estimate & $95\%$ Credible Interval & Estimate & $95\%$ Credible
Interval\\ \midrule
$\beta_{\text{Age}}$ 
& 0.1847 & (0.1693, 0.2158) & 0.0728 & (0.0138, 0.2056)\\
$\beta_{\text{Sex}}$ 
& 0.1239 & (0.0912, 0.1732) & 0.3411 & (0.1118, 0.5189) \\ 
$\beta_{\text{Grade}}$ 
& 0.5075 & (0.4366, 0.5291) & 0.8290 & (0.4583, 0.9578) \\ 
$\beta_{\text{Hist-Stage}}$ 
& 1.3271 & (1.2926, 1.3824) & 1.4434 & (1.3150, 1.7101) \\ 
$\lambda_1$ & 1.0690 & (0.9938, 1.0912) & 1.0359 & (0.9060, 1.2330) \\ 
$\lambda_2$ & 1.0716 & (0.9909, 1.1172) & 1.0877 & (0.8432, 1.3035) \\ 
$\lambda_3$ & 0.9843 & (0.9787, 1.1040) & 1.0912 & (0.8535, 1.2899) \\ 
$\lambda_4$ & 1.0040 & (0.9583, 1.0513) & 1.0209 & (0.8724, 1.2091) \\ 
\bottomrule
\end{tabular}
\end{table}

\section{Discussion}\label{sec:discussion}
In this paper, we proposed a geographically weighted Chinese restaurant 
process to capture spatial homogeneity of regression coefficients and 
baseline hazards based on piecewise constant hazard model. An efficient MCMC
algorithm is proposed for our methods without complicated reversible jump 
algorithm. Extensive simulation results are carried out to show that 
our proposed method has better clustering performance than the traditional 
CRP in spatial homogeneity pursuit for survival data. Simulation studies also 
show that our proposed methods have promising results in coefficients and 
baseline hazard estimation. An application to analysis of SEER data provides 
an interesting illustration of our proposed methods.

Furthermore, four topics beyond the scope of this paper are worth further
investigation. In this paper, our proposed
algorithm is based on two step estimation under piecewise constant 
proportional hazard model assumption. Proposing an efficient
sampling algorithm without Laplace approximation is an important future work.
Furthermore, we fixed the number of pieces of baseline hazards in both 
simulation studies and real data analysis. Imposing adaptive number of pieces 
model in baseline hazards is devoted for future research.
In addition, variable selection approaches based on hierarchical CRP 
\citep{griffiths2004hierarchical} is also worth being investigated. Allowing
different covariates and baseline hazard share
different clustering processes is also an important future work.

\section*{Acknowledgement}
The authors would like to thank the editor, the associate editors and two
reviewers for their valuable comments which help improve the presentation of
this paper.

\section*{Data Availability Statement}
The data that support the findings of this paper are available from the
corresponding author upon reasonable request.

\bibliographystyle{chicago}
\bibliography{spatial_survival}

\section*{Supporting Information}

Web Appendices, Tables, and Figures referenced in Sections 2-5 and R scripts
for simulationsand real data examples are available with this paper at the
Biometrics website
on WileyOnline Library. The R code
for the computations of this paper is available at
\url{https://github.com/lj-geng/GWCRP}.
\end{document}